\documentclass[prl,showpacs,twocolumn]{revtex4}
\usepackage{graphicx}
\begin{document}

\title{Mott insulators in an optical lattice with high filling factors.}
\author{D. van Oosten,$^{1,2}$ P. van der Straten,$^2$ and
        H. T. C. Stoof $^1$}
\affiliation{$^1$Institute for Theoretical Physics,
         Utrecht University, Leuvenlaan 4, 3584 CE Utrecht, The Netherlands\\
$^2$Debye Institute,
         Utrecht University, Princetonplein 5, 3584 CC Utrecht, The Netherlands}

\begin{abstract}
We discuss the superfluid to Mott insulator transition 
of an atomic Bose gas in an optical lattice with high filling factors.
We show that also in this multi-band situation, the long-wavelength physics is described 
by a single-band Bose-Hubbard model. We determine the 
many-body renormalization of the tunneling and interaction parameters in the effective
Bose-Hubbard Hamiltonian,
and consider the resulting model at nonzero temperatures. We show that in particular for
a one or two-dimensional optical lattice, the Mott insulator phase
is more difficult to realize than anticipated
previously.
\end{abstract}

\maketitle

{\it Introduction.}--- The behaviour of trapped Bose-Einstein
condensates offers a large amount of interesting features.
Specifically, the phase coherence of a condensate creates the
prospect of various interference experiments, as shown for the first time 
in an experiment performed by Andrews {\it et al.}
\cite{wolfgang}. In another experiment  a large number of condensates trapped in
a periodic lattice potential have been made to interfere
\cite{mark}. Having shown experimentally that in general condensates are
phase coherent and thus show off-diagonal long-range order,
the question arises if this long-range order can also
be destroyed in a controllable way. This was only very recently
achieved in a beautiful experiment by Greiner {\it et al.}
\cite{Bloch}.

In this last experiment, a trapped Bose-Einstein condensate is
put into a three-dimensional optical lattice. The number of
atoms in the condensate is sufficiently large to obtain a filling
factor of almost two atoms per site. By increasing the intensity of the
lattice light, a quantum phase
transition from a superfluid state to a Mott insulating state is
achieved. In the insulating phase all phase coherence is lost due
to quantum fluctuations. The transition was predicted to occur in
this system by Jaksch {\it et al.} \cite{Jaksch}, and the observed
critical conditions for the transition are in good agreement with
the results of an on-site mean-field theory \cite{Oosten}. 
This indicates that the single-band Bose-Hubbard model 
used in Refs.~\cite{Jaksch,Oosten} can accurately describe a gas
of ultracold bosonic atoms in an optical lattice when the filling
factor of the lattice is of the order of one. 

However, this model is no longer valid in the case of higher filling
factors such as described in the experiments of \cite{Kasevich,Esslinger}.
The theories mentioned above use single-particle wave functions corresponding
to the lowest band of the lattice to calculate the microscopic parameters
of the single-band Bose-Hubbard model as a function of the lattice parameters.
In the case of high filling factors more than one band is generally 
populated, leading to a multi-band Bose-Hubbard model.
The interaction effects that occur under these circumstances have not been considered previously.
Furthermore, the effects of thermal fluctuations are also not understood,
even in the single-band Bose-Hubbard model. There are studies that describe number
squeezing in an optical lattice at nonzero temperature \cite{Juha,Burnett},
but they are not in the strongly-interacting limit that is relevant for the Mott
insulator.
Our main objective here is to develop an effective theory that can deal with 
these issues.

{\it Solving the multi-band Bose-Hubbard model.} --- To
solve the problems associated with
high filling factors, we have to deal with the many-body physics at every site.
Since the high filling factors of interest are experimentally most relevant 
in low-dimensional lattices \cite{Kasevich,Esslinger}, we  
discuss the energy scales involved in those systems. In a 
low-dimensional lattice, we can approximate the on-site trapping
potential by an anisotropic harmonic potential. The oscillator 
frequencies $\omega_\parallel$ and
$\omega_\perp$ correspond to the trapping frequencies in the
directions parallel and perpendicular to the periodicity of the
lattice, respectively. Because the typical size of a lattice well
in the parallel direction(s) is much smaller than in the
perpendicular direction(s), we immediately have that 
$\hbar\omega_\perp \ll \hbar \omega_\parallel$. Furthermore, for the
experimental conditions of interest \cite{Kasevich,Esslinger}, the
temperature is in between the two trapping frequencies, {\it i.e.},
$\hbar \omega_\perp \ll k_B T \ll \hbar \omega_\parallel$. This
implies that in every site the gas is in the parallel direction(s) in the ground
state of the potential, but that it occupies many states in the
perpendicular direction(s). In particular, this holds for the
thermal cloud of the gas. As a consequence, the effective
dimensionality of the gas at every site is reduced 
and the thermal excitations are only present in the perpendicular
direction(s). It is under these conditions that we are able to solve 
the relevant multi-band Bose-Hubbard model by using the following two-step
procedure.

First, we solve the many-body physics at every site. 
Due to the famous infrared problems of
a one or two-dimensional Bose gas this is not an easy task, and an accurate
equation of state for these gases was developed only very recently
in the weakly-interacting limit \cite{stoof}. 
This equation of state is found by treating phase fluctuations in the
(quasi)condensate exactly and we can in particular use it to
determine at every temperature the number of atoms in the
(quasi)condensate $N_0(T)$. Furthermore, it is shown in Ref.~\cite{stoof} that
even in the presence of phase fluctuations, the Gross-Pitaevskii
equation can still be used to calculate the density profile of the (quasi)condensate.

Secondly, we consider the coupling between the sites.
Since we have a (quasi)condensate at every site, the coupling between
sites will be dominated by tunneling from (quasi)condensate to (quasi)condensate
as opposed to (quasi)condensate to thermal cloud. This means that 
we can describe the coupling between sites by a single-band
Bose-Hubbard model. The important parameters in 
the Bose-Hubbard model are the on-site
interaction energy $U$ and the energy $t$ associated with the
tunneling of atoms between nearest-neighbor sites. Both energies
can be calculated from the knowledge of the (quasi)condensate wave
function $\psi_0({\bf x})=\sqrt{n_0({\bf x})} e^{i\vartheta}$,
where $n_0({\bf x})$ is the density profile and $\vartheta$ is the
global phase of the (quasi)condensate. The interaction energy $U$
is proportional to $\int d{\bf x} |\psi_0({\bf x})|^4$, whereas
the tunneling energy $t$ requires the evaluation of an overlap
integral between the (quasi)condensate wave functions of two
neighboring sites in the parallel direction(s).

To describe the effect of the interatomic interaction, we thus
need to determine how the (quasi)condensate wave function changes
as a result of the on-site interactions. 
Since the mean-field interaction obeys 
$\hbar \omega_{\perp} \ll N_0 U \ll \hbar\omega_{\parallel}$
under the experimental conditions of interest, we can write the 
three-dimensional wave function of the condensate as a product
of a single-particle ground-state wave function in the parallel
direction(s) and the (quasi)condensate wave function in the
perpendicular direction(s). If we substitute this product wave
function into the Gross-Pitaevskii equation and integrate out
the parallel direction(s), we arrive at an effective 
equation for the (quasi)condensate wave function. Because of the
above mentioned inequality, we can subsequently solve this
equation using the Thomas-Fermi or local-density approximation \cite{Dalfovo}.
To quantify the differences between the (quasi)condensate wave function
and the single-particle ground-state wave function, we
define a dimensionless coupling constant $g$ both in the noninteracting and in
the interacting case. The first parameter we call the bare coupling
constant $g_{\rm B}=U_{\rm B}/t_{\rm B}$ and it is calculated with the single-particle 
ground-state wave function in every site. The second parameter we call the
renormalized coupling constant $g_{\rm R}=U_{\rm R}/t_{\rm R}$ and it is calculated using the
single-particle ground state in the parallel direction(s) and a
Thomas-Fermi density profile in the perpendicular direction(s).
Because we have already included the on-site interaction effects in this coupling 
constant, we can now write down a renormalized single-band Bose-Hubbard model
for the total optical lattice, 
where the creation and annihilation operators ${\hat a}_i^{\dagger}$, and ${\hat a}_i$, respectively,
and  the number operator ${\hat n}_i$ are not associated with the Wannier states of atoms in the lattice,
but with the macroscopic wave function of the (quasi)condensate in each site. In particular,
we have

\begin{eqnarray}
{\hat H}&=& - t_{\rm R} \sum_{\langle i,j \rangle} {\hat a}_i^\dagger 
{\hat a}_j 
+\frac{U_{\rm R}}{2} \sum_i {\hat n}_i \left({\hat n}_i -1 \right) -\mu_{\rm R} \sum_i {\hat n}_i,
\label{eq:renormBHM}
\end{eqnarray}
where $\mu_{\rm R}$ is the effective chemical potential. 
The interaction parameter is given by
$U_{\rm R}=
\left.\partial^2 F^{\rm os}/\partial N^2\right|_{N=N_0}
\equiv\left.\partial \mu^{\rm os} /\partial N \right|_{N=N_0}$, where $F^{\rm os}$ is the
on-site free energy and $\mu^{\rm os}$ is 
the on-site chemical potential. Formally, the effective chemical potential is given by
$\mu_{\rm R}=\mu - \mu^{\rm os} - U_{\rm R}/2$, where the last term is substracted 
from the interaction energy to make the analogy to
the single-band Bose-Hubbard model of Refs.~\cite{Jaksch,Oosten} complete.

It is important to understand that the hopping term
only describes hopping between the (quasi)condensates 
in neighbouring sites. While it is clear that this is a very good approximation in 
the case of neighbouring condensates, it may not be immediately obvious in the 
case of neighbouring quasicondensates. However, it should be noted that
the effect of the hopping is only large when
the system is in the superfluid phase, in which case all the sites couple to
form a true three-dimensional condensate.
The tunneling strength can be calculated in the tight-binding
limit and depends only on the overlap in the parallel direction(s).
As a result the bare and renormalized values of $t$ are equal.
However, the interaction energy is strongly reduced due to
the repulsive on-site interactions which spread-out the (quasi)condensate
wave function considerably. 
We find for a two-dimensional gas that
\begin{eqnarray}
g_{\rm R} &=& g_{\rm B} \left( \frac{\pi}{2} \right)^{1/4}
\left(\frac{\ell_\parallel}{N_0 a}\right)^{1/2} \propto g_{\rm B}
\left( \frac{\ell_\perp}{R_{\rm TF}} \right)^2,
\label{eq:renorm2D}
\end{eqnarray}
and for a one-dimensional gas that
\begin{eqnarray}
g_{\rm R} &=& g_{\rm B} 
\left( \frac{\pi}{2} \right)^{1/2} 
\left(\frac{\ell_\parallel^2}{3 N_0 a \ell_\perp}\right)^{1/3}
\propto g_{\rm B} \left( \frac{\ell_\perp}{R_{\rm TF}} \right).
\label{eq:renorm1D}
\end{eqnarray}
Here $a$ is the positive $s$-wave scattering length of the atoms,
$\ell_\parallel=\sqrt{\hbar/m\omega_\parallel}$ and
$\ell_\perp=\sqrt{\hbar/m\omega_\perp}$ are the harmonic
oscillator lengths in the parallel and perpendicular directions,
respectively, and $R_{\rm TF}$ is the Thomas-Fermi radius of the
(quasi)condensate. The physical interpretation of
Eqs.~(\ref{eq:renorm2D}) and (\ref{eq:renorm1D}) is that 
as a result of the repulsive interatomic interactions, the (quasi)condensate
reduces its total energy by increasing its size in the perpendicular direction(s).
This can be seen from the fact that
the decrease of the coupling constant is inversely proportional to
the increase in the surface or length of the two or
one-dimensional gas, respectively.
Note that his reduction is particularly important for the critical conditions,
which can be written as $g_{\rm R}>4zN_0$ for large $N_0$ \cite{Oosten}.
In order to verify the consistency of our two-step approach, we explicitly
check the relevant energy scales using our results of Eqs.(\ref{eq:renorm2D}) and 
(\ref{eq:renorm1D}). First, we assumed
that the mean-field interaction energy $N_0 U_{\rm R}$ is much
smaller than the trapping frequency in the parallel direction(s)
$\hbar\omega_\parallel$. This requires that for a one-dimensional
lattice
\begin{eqnarray}
N_0 \ll \left( \frac{\hbar \omega_\parallel}{\hbar\omega_\perp}
\right)^2 \sqrt{2 \pi} \frac{\ell_\parallel}{a}.
\label{eq:checkmf}
\end{eqnarray}
Second, we also assumed that the crossover temperature for the
formation of a (quasi)condensate in two dimensions is much lower
than $\hbar\omega_\parallel$. This results in
\begin{eqnarray}
\hbar \omega_\perp \left( \frac{N}{\zeta (2)} \right)^{1/2} \ll
\hbar \omega_\parallel, \label{eq:checktc}
\end{eqnarray}
where $N$ is the total number of atoms at every site. For typical
numbers used in the experiments by Orzel {\it et
al.}~\cite{Kasevich}, we find the condition $N_0 \leq N \ll 10^5
$, which means that our assumptions are valid for even the largest filling factor reported.
For the case of the two-dimensional lattice of
Greiner~{\it et al.} \cite{Esslinger}, a similar
inequality can be derived. It is found that this experiment is also in the regime where
our assumptions are valid. Note that the use of the Thomas-Fermi
approximation also imposes a lower limit on the filling factor,
namely, $N_0 a \gg l_\perp$. However, when the filling factor is 
below this limit, we are in the regime where we can safely use the bare
coupling constant.
We thus conclude that depending on the filling factor, either our renormalized
or the bare theory is applicable to these experiments.

{\it Thermal effects.} --- Besides the effect of thermal
fluctuations on the number of (quasi)condensate atoms per site
$N_0(T)$, which is accounted for by the equation of state of the
low-dimensional Bose gas \cite{stoof}, there is also the effect of thermal
fluctuations on the renormalized single-band Bose-Hubbard model itself.
These thermal fluctuations are also present in a lattice with low filling factor.
To study these thermal fluctuations, we use an on-site mean-field Hamiltonian that we
can derive from Eq.~(\ref{eq:renormBHM}),
using the approach presented in Ref.~\cite{Oosten}.
We decouple the tunneling term, by introducing a complex mean field parameter 
$\psi$ as follows $\hat{a}_i^\dagger\hat{a}_j=\psi \hat{a}_j + \hat{a}_i^\dagger
\psi^* - \left| \psi \right|^2$.
Physically, $\psi$ is the  superfluid order parameter which we choose to be real
in the following. Performing the above substitution, we find

\begin{equation}
{\hat H}(\psi)=  -z t\psi({\hat a}^\dagger+{\hat a})
    +\frac{U_{\rm R}}{2} {\hat n}({\hat n}-1)
    -\mu_{\rm R} {\hat n}
    + z t \psi^2,
\label{eq:hamiltonian}
\end{equation}
with $z$ the coordination number. Since this is an on-site Hamiltonian,
we have dropped the site indices for simplicity. 
Moreover, we assumed that the chemical potential is chosen such that the 
expectation value of the number operator $\hat{n}$ is equal to the number
of quasi(condensate) particles $N_0(T)$ in every site.

\begin{figure}[t]
\includegraphics[width=7cm]{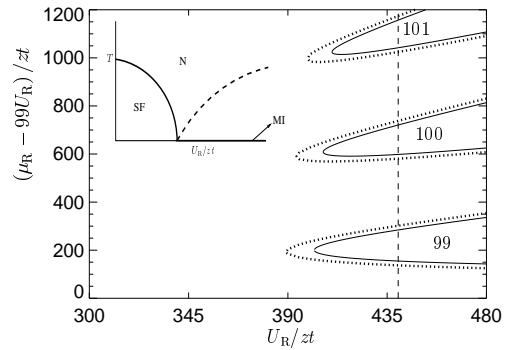}
\caption{Phase diagram  of the Bose Hubbard model 
in terms of the chemical potential $\mu_{\rm R}/zt$ and the coupling constant
$U_{\rm R}/zt$. The solid and dotted lines correspond to $T=0$ and $T=0.1U_c$ (where $U_c$ is the 
critical $U_{\rm R}$ for the $N_0=100$ lobe), respectively.
The inset shows a
qualitative phase diagram in terms of the temperature $T$ and the
coupling constant $g$. N, SF and MI indicate the normal gas phase,
the superfluid and the Mott insulating phase, respectively.}
\label{fig:phasediag}
\end{figure}

\begin{figure}[t]
\includegraphics[width=7cm]{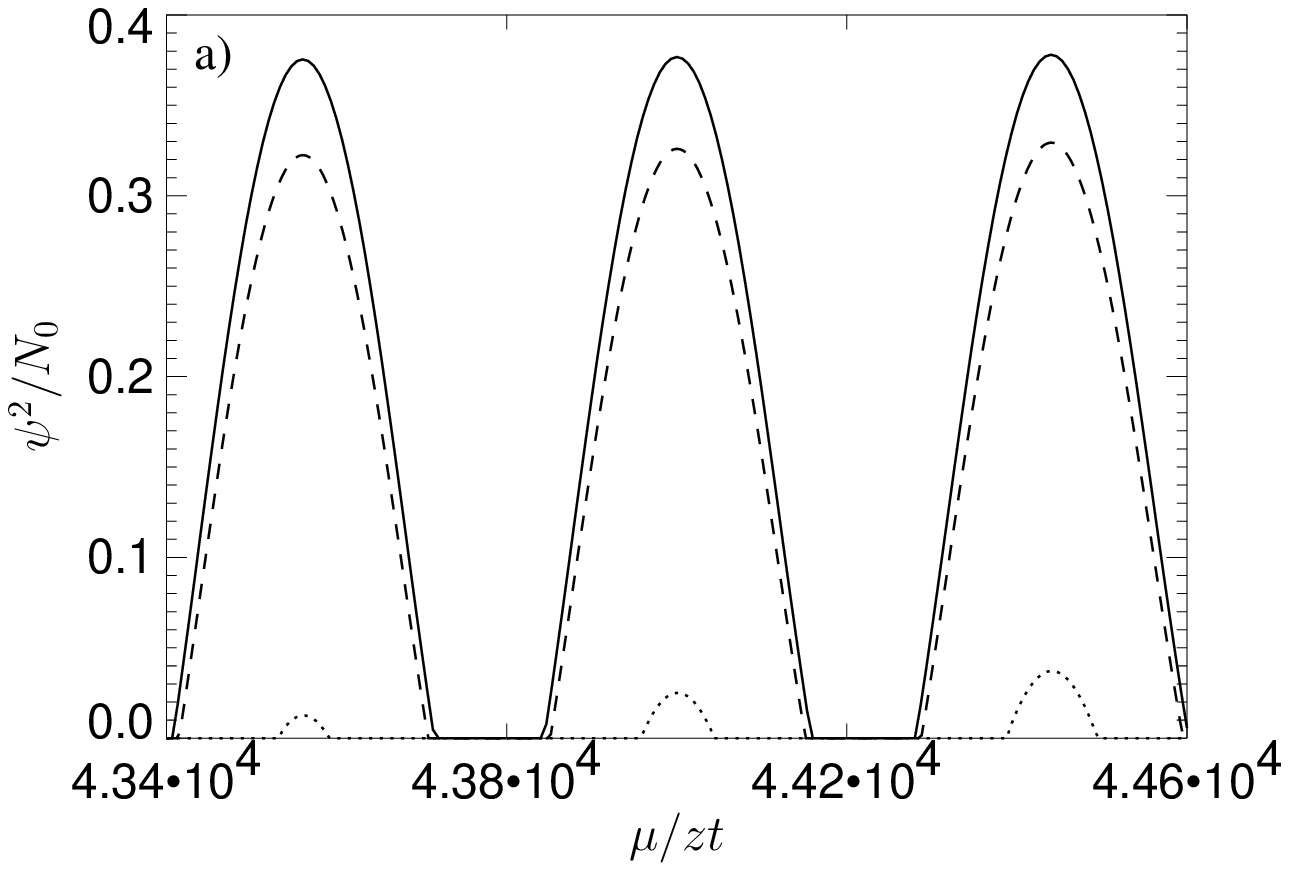}
\includegraphics[width=7cm]{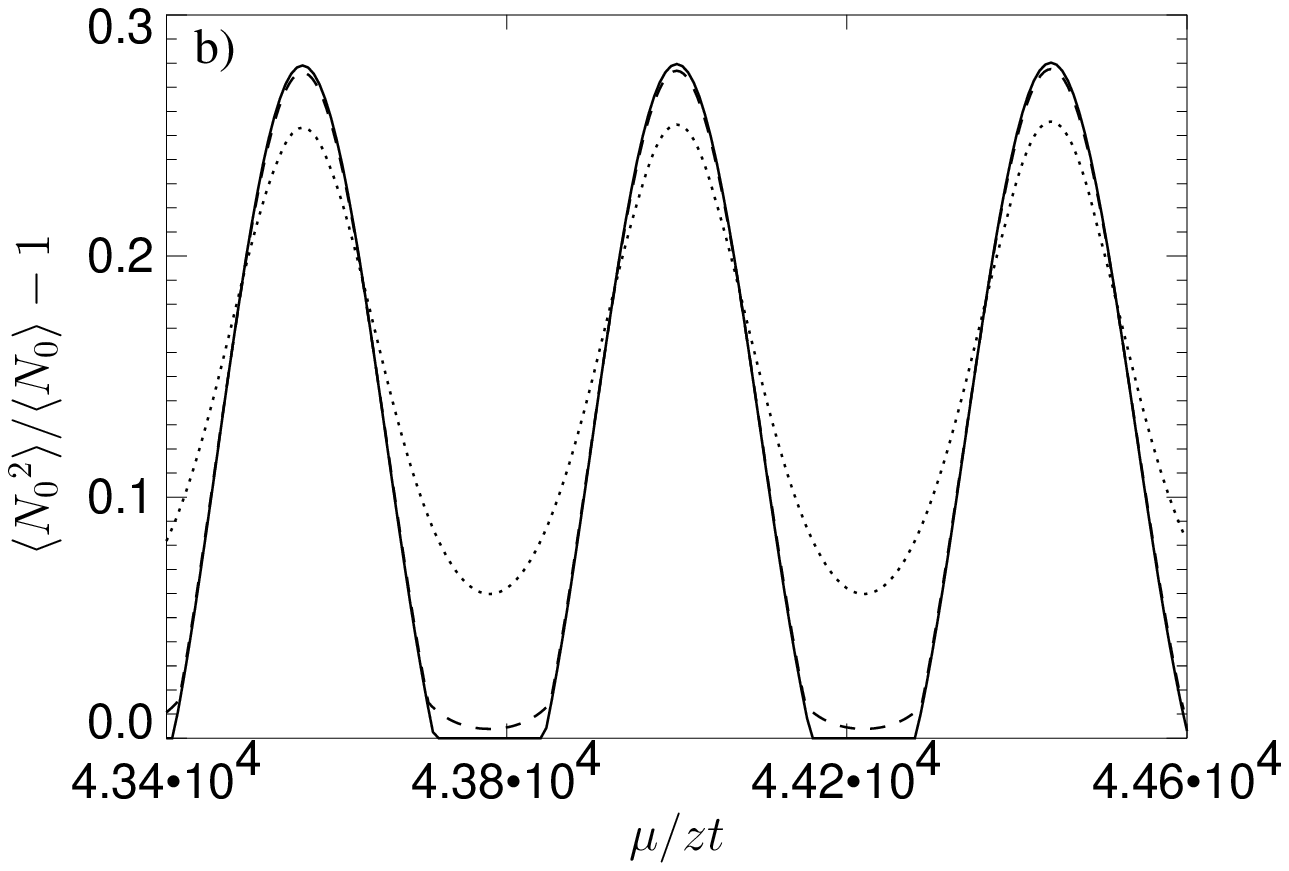}
\caption{Condensate fraction (a)
and particle number fluctuations (b) as a function of the coupling
constant, for $k_B T = 0 , 0.08 U$ and $0.145 U$ (solid, dashed and dotted lines respectively).
The coupling constant $U_{\rm R}/zt=440$ (the dashed line in Fig~\ref{fig:phasediag}).}
\label{fig:thermal}
\end{figure}

The zero-temperature phase diagram of this mean-field theory can
be solved exactly \cite{Oosten,Sachdev} and is shown in
Fig.~\ref{fig:phasediag}, where the Mott insulator
phases correspond to the various lobes. 
For nonzero temperatures the model can no longer be solved analytically
and we have to resort to numerical methods. 
If we put $\psi=0$, we find
that the eigenstates of Eq.~(\ref{eq:hamiltonian}) are given by number
states. Using a basis consisting of these number states, running from
a certain minimum filling factor $N_{\rm min}$ to a certain maximum $N_{\rm max}$,
we can calculate the grand-canonical
partition function $Z(\psi)={\rm Tr}[e^{-H(\psi)/k_B T}]$ by diagonalizing the 
mean-field Hamiltonian given in Eq.~(\ref{eq:hamiltonian}).
Next we determine  
the thermodynamic potential $\Omega(\psi)=-k_{\rm B} T \ln{Z(\psi)}$ as a function of the
order parameter $\psi$. For zero temperature the calculation converges when 
$N_{\rm max} \simeq N_0 + 4$ and 
$N_{\rm min} \simeq N_0 - 4$, where $N_0$ is the 
filling factor of the relevant Mott-insulator lobe. For nonzero
temperatures, more states must be included.

To obtain the relevant thermodynamic quantities, we minimize the grand
potential $\Omega(\psi)$ and the value of $\psi$ at the minimum of $\Omega(\psi)$
corresponds physically to the square root of the number of atoms that is superfluid in
the direction(s) parallel to the periodicity of the lattice. 
In the Mott insulator the gas is only insulating in the direction(s)
parallel to the periodicity of the optical lattice, whereas it is always a superfluid
in the perpendicular direction(s).
The other quantity of interest is the value of the number fluctuations.
This number is important because in the Mott-insulator phase the number
fluctuations are exactly zero. 

The final results of the calculations are shown in Fig.~\ref{fig:thermal}.
In these figures,
the longitudional superfluid fraction and the number fluctuations are plotted along the
dashed line in Fig.~\ref{fig:phasediag} for different temperatures.
It can clearly be
seen from Fig.~\ref{fig:thermal}(a) that the superfluid part of
the phase diagram decreases with increasing temperature.
In addition Fig.~\ref{fig:thermal}(b) shows that at zero temperature the
density fluctuations drop exactly to zero in the Mott insulating
regions, but that this does not happen at nonzero temperature.
This is a result of the fact that
the superfluid to Mott insulator transition is a quantum
phase transition. The reason that there is still a reduction in particle-number
fluctuations at nonzero temperature is that the excitation spectrum 
of a fluctuation is gapped in this region \cite{Oosten}, which
means that the fluctuations are exponentially suppressed. 
Due to this strong suppression of the number fluctuations, 
one will be able to observe a phase which is formally not a Mott insulator,
but experimentally has very similar features. Another feature we can clearly see in
Fig.~\ref{fig:thermal}(b) is that part of the phase diagram where the number 
fluctuations are suppressed also decreases with increasing temperature,
and shrinks in the opposite direction of that of the superfluid part. 

On the basis of the above calculations, we can draw the nonzero temperature 
phase diagram shown in Fig.~\ref{fig:phasediag}.
In this figure, the solid lines indicated the
superfluid to Mott insulator transition at zero temperature and the dotted
lines indicate the superfluid to normal transition at nonzero temperature.
The inset shows the phase diagram in terms
of the temperature and the coupling constant. This diagram agrees very well with
the general description given by Sachdev \cite{Sachdev}.

{\it Conclusion} - We have shown that for low-dimensional lattices,
which generally have a filling factor much larger than one, we should in principle 
solve a many-band Bose-Hubbard model. This can be achieved
by first solving the on-site many-body problem, and then deriving an effective theory
that describes the coupling between the sites in the optical lattice in terms of a
renormalized single-band Hubbard model.
We have calculated the effects of thermal excitations in this 
renormalized model and we have shown that the number fluctuations in the above model can
only drop to zero in the absence of thermal fluctuations. However, if the temperature
is sufficiently low, the number fluctuations are exponentially suppressed. 
This means that at a certain nonzero temperature, the crossover to the Mott
insulator phase can still be observed if the coupling constant is increased to a
value larger than the zero-temperature critical value (cf. Fig.~\ref{fig:phasediag}).
It is important to realize that to experimentally obtain the Mott insulator
with a large filling factor $N_0$, the coupling constant
$g_{\rm R}=U_{\rm R}/t$ must be larger than  $ 4 z N_0 $.
However, Eqs.~(\ref{eq:renorm2D}) and (\ref{eq:renorm1D}) show that
the renormalized coupling constant is much smaller than the bare coupling constant
for a low-dimensional optical lattice.
We therefore conclude that in terms of the bare coupling
constant, which is the experimentally relevant control parameter,
the Mott insulator phase is much more difficult to obtain than is
naively anticipated on the basis of a purely single-band Bose-Hubbard
model.


\begin{thebibliography}{99}
\bibitem{wolfgang}
    M. R. Andrews, C. G. Townsend, H. -J. Miesner, D. S. Durfee,
    D. M. Kurn, and W. Ketterle,
    Science {\bf 275}, 637 (1997).
\bibitem{mark}
    B. P. Anderson, M. A. Kasevich,
    Science {\bf 282}, 1686 (1998).
\bibitem{Bloch}
    M. Greiner, O. Mandel, T. Esslinger, T. W. H\"ansch, and I. Bloch,
    Nature {\bf 415}, 39 (2002).
\bibitem{Jaksch}
    D. Jaksch, C. Bruder, J. I. Cirac, C. W. Gardiner and P. Zoller,
    Phys. Rev. Lett. {\bf 81}, 3108 (1998).
\bibitem{Oosten}
    D. van Oosten, P. van der Straten, and H. T. C. Stoof,
    Phys. Rev. A {\bf 63}, 053601 (2001).
\bibitem{Kasevich}
    C. Orzel, A. K. Tuchman, M. L. Fenselau, M. Yasuda, and M. A. Kasevich,
    Science {\bf 291}, 2386 (2001).
\bibitem{Esslinger}
    M. Greiner, I. Bloch, O. Mandel, T. W. H\"ansch, and T. Esslinger,
    Phys. Rev. Let. {\bf 87}, 160405 (2001).
\bibitem{Juha}
    J. Javanainen, Phys. Rev. A {\bf 60}, 4902 (1999).
\bibitem{Burnett}
    K. Burnett, M. Edwards, C. W. Clark, and M. Shotter,
    J. Phys. B {\bf 35}, 1671 (2002).
\bibitem{Sachdev}
    S. Sachdev, {\it Quantum Phase Transitions} (Cambridge Univ. Press, Cambridge, 2001).
\bibitem{stoof}
    J. O. Andersen, U. Al Khawaja, and H. T. C. Stoof,
    Phys. Rev. Lett. {\bf 88}, 070407 (2002);
    U. Al Khawaja, J. O. Andersen, N. P. Proukakis, and H. T. C. Stoof,
    Phys. Rev. A. {\bf 66}, 013616 (2002).
\bibitem{Dalfovo}
	F. Dalfovo, S. Giorgini, and L. P. Pitaevskii, Rev. Mod. Phys.
	{\bf 71}, 463 (1999).
\end{thebibliography}
\end{document}